| | |
|---|---|
| Title | **Optical diagnostics and mass spectrometry on the afterglow of an atmospheric pressure Ar/O$_2$ radiofrequency plasma used for polymer surface treatment** |
| Authors | C. Y. Duluard, T. Dufour, J. Hubert, F. Reniers |
| Affiliations | Université Libre de Bruxelles, Faculté des Sciences, Chimie Analytique et Chimie des Interfaces, Boulevard du Triomphe,2 – 1050 Bruxelles, Belgium |
| Ref. | 30th-ICPIG, Belfast, Northern Ireland, 28th August – 2nd September 2011, C10 http://mpserver.pst.qub.ac.uk/sites/icpig2011/301_C10_Duluard.pdf |
| DOI | - |
| Abstract | In the context of polymer surface treatment, the afterglow of an atmospheric pressure Ar/O$_2$ radiofrequency plasma is characterized by optical emission spectroscopy, laser induced fluorescence and mass spectrometry. The influence of the O$_2$ gas flow rate and the source power on the plasma properties (gas temperature, Ar excitation temperature, relative concentrations of O atoms and OH radicals) are evaluated. We show that for plasma torch-to-substrate distances lower than 6 mm, the afterglow creates a protective atmosphere, thus the plasma gas composition interacting with the substrate is well controlled. For higher distances, the influence of ambient air can no longer be neglected and gradients in Ar, O$_2$ and N$_2$ concentrations are measured as a function of axial and vertical position. |

# 1. Introduction

Non-equilibrium atmospheric pressure plasmas are gaining an interest for thin film deposition, surface treatment and biomedical applications for their ease of handling and their low operating gas temperatures. The atmospheric pressure RF plasma torch studied for this paper is an Atomflo™-250D plasma source from Surfx Technologies LLC, running at a frequency of 27.12 MHz, and commercialized for plasma enhanced chemical vapour deposition or for treating polymer surfaces to increase, for instance, their adhesion properties. Helium is normally the main plasma gas, but this has been replaced with argon in our laboratory. With the ever increasing rarity and expected rises in the price of helium, the choice to use argon has been consciously made to prepare for the future scenario where helium becomes less commercially viable. The discharge is produced between two multi-perforated, parallel plate electrodes, 2.5 cm in diameter. The Ar and O$_2$ gas flows range between 20-40 L/min and 0-30 mL/min respectively. The power injected in the discharge is set between 60 W and 100 W. The polymer samples to be treated are placed downstream of the two electrodes, at a distance varying from 2 mm to 10 mm. This work aims to characterize by optical emission spectroscopy, laser induced fluorescence and mass spectrometry the Ar/O$_2$ plasma afterglow interacting with polymer surfaces, in order to bring to light the mechanisms responsible for their surface modification.

# 2. Characterisation of Ar/$O_2$ plasma afterglow: influence of source power and $O_2$ gas flow

## 2.1. OH radicals, O atoms

One species of interest, though probably in minor concentration, is the OH radical, which presents a high oxidative reactivity and may participate in the surface activation of polymers such as LDPE (low density polyethylene). It is easily detected in our conditions by optical emission spectroscopy through the $A^2\Sigma^+, v'=0 \rightarrow X^2\Pi, v''=0$ emission band around 309 nm. Ar and $O_2$ gases purities are 99.999%, and several tests (including experiments presented in section 2) lead us to conclude that the OH production mostly comes from water that adsorbs on the inner walls of the plasma torch when the plasma is turned off. However, it seems that the radical concentration decays as the plasma on time increases, see figure 1. For all the experiments presented, water vapour was then injected under the electrodes to allow a better stabilization of the OH concentration.

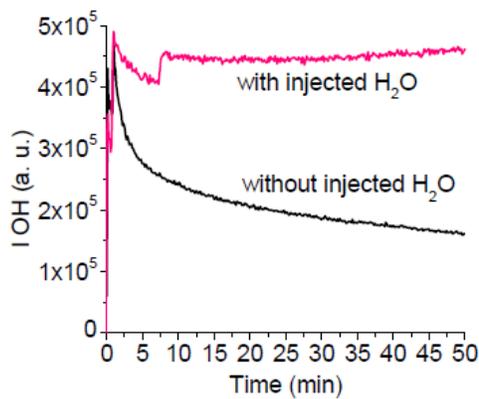

*Figure 1 OH ($A^2\Sigma^+, v'=0 \rightarrow X^2\Pi, v''=0$) band emission intensity versus plasma on time, with and without injected water vapour (Ar 30 L/min, 80 W, plasma torch-to-substrate distance 5 mm)*

Laser induced fluorescence experiments were carried out to measure the relative OH concentration as a function of plasma parameters. The laser pumping was set on the P2(6) transition at 284.9 nm which excites the F2(5) level of the upper OH($A^2\Sigma^+$, v'=1) vibrational state from the F2(6) level of the ground OH($X^2\Pi$, v''=0) vibrational state [1]. The laser pumped level can deexcite through several mechanisms: fluorescence, collisional quenching down to the fundamental, and rotational (RET) and vibrational energy transfer (VET). VET down to the OH ($A^2\Sigma^+$, v'=0) level can be problematic since the resulting emission in the (0,0) band overlaps the (1,1) band emission measured. Simulation of energy transfers with the LASKIN program [2] indicated that the (0,0) band emission following VET was at most one order of magnitude lower than the (1,1) band emission for $N_2$ and $CO_2$ concentrations of 0.01%, which is the case for small plasma torch-to-substrate distances (cf. section 2). Taking into account the variations in decay time due to collisional quenching, we could then obtain trends of the OH concentration in the fundamental state. Figure 2 shows the ratio of the integrated LIF signal to the fluorescence decay time, considered to be proportional to the OH concentration, versus plasma source power. Globally, the OH concentration decreases with an increase in plasma source power and is higher when $O_2$ is injected with Ar.

The major oxidative species expected is atomic oxygen [3]. OES spectra show a linear increase in excited state $O_{3p}$ density with the increase in $O_2$ gas flow (cf. figure 3). Ideally, two-photon LIF measurements on the O radical could be undertaken to verify that the ground state density follows indeed this trend, because emission is greatly dependent on the electron energy distribution function.

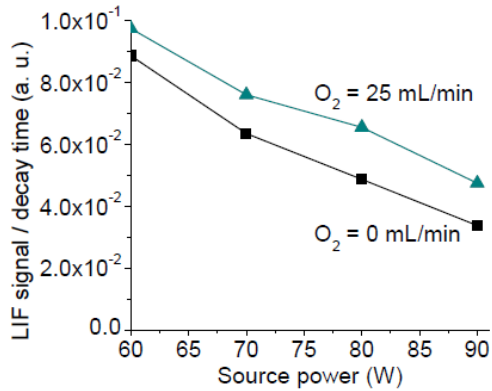

*Figure 2 Relative OH concentration versus plasma source power (Ar 30 L/min, plasma torch-to-substrate distance 5 mm)*

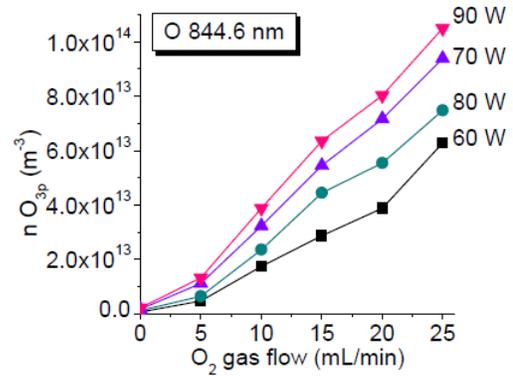

*Figure 3 $O_{3p}$ density from measurement of the O 844.6 nm line intensity versus $O_2$ gas flow*

## 2.2. Temperature measurements

An Ar excitation temperature $T_{13}$ was derived from the calculation of the Ar ground state density (level 1) and from the determination of the Ar 4p levels density (level 3) via absolute optical emission spectroscopy. Because of the energy difference between the ground state and the 4p levels (12.5-13.5 eV), this method is more accurate to obtain an excitation temperature in non-equilibrium plasmas [4]. A collisional radiative model is then needed to convert the estimated temperature $T_{13}$ to an electron temperature. In this study, only variations in $T_{13}$ with the plasma parameters are presented.

The Ar ground state density was derived from the ideal gas law, $P=n_1kT_{gas}$, where P is 1 bar and $T_{gas}$ was estimated by the OH rotational temperature. This equation was used assuming that the population of the Ar ground state was much greater than that of any other species in the plasma afterglow. To obtain the OH rotational temperature, the OH (0,0) band emission was fitted to simulated spectra using the software LIFBASE [5]. In all the conditions tested, $T_{rot}$ was found in the 400-440 K range.

Figure 4 shows that $T_{13}$ increases with $O_2$ gas flow when greater than 5 mL/min, and increases with source power. It is thus possible that the O ground state density does not show a linear increase with the $O_2$ gas flow as suggested by OES measurements.

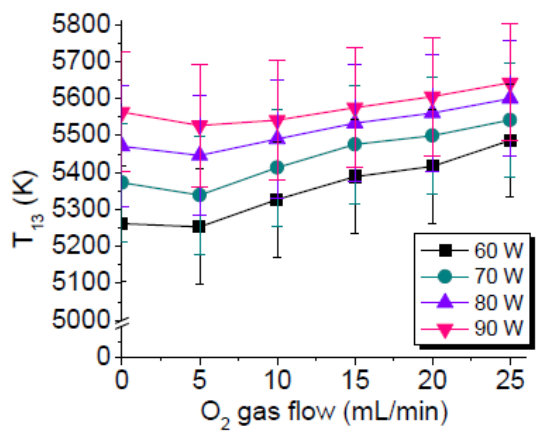

*Figure 4 Ar excitation temperature $T_{13}$ versus $O_2$ gas flow for various powers (Ar 30 L/min, plasma torch-to-substrate distance 5 mm)*

## 3. Influence of plasma torch-to-substrate distance

The plasma torch-to-substrate distance greatly affects the plasma afterglow properties. Figure 5 shows that the Ar excitation temperature $T_{13}$ decreases as the substrate is moved away from the plasma torch.

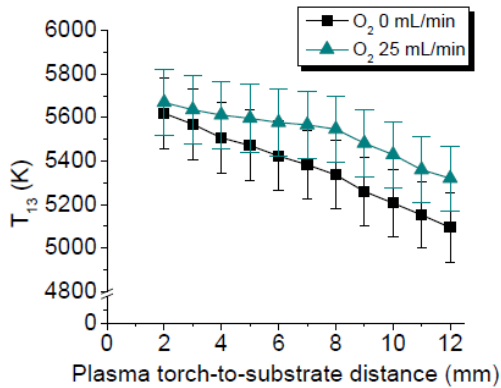

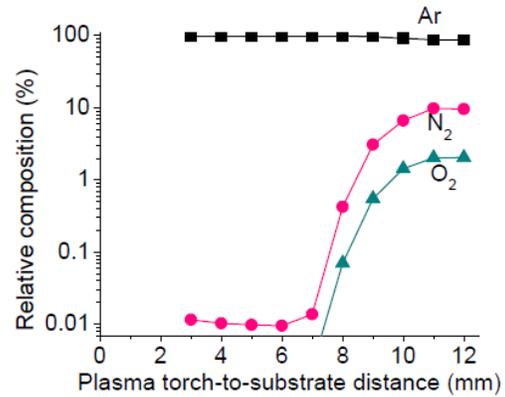

*Figure 5 Ar excitation temperature $T_{13}$ versus plasma torch-to-substrate distance (Ar 30 L/min, 80 W)*

*Figure 6 Ar, $N_2$ and $O_2$ relative concentrations determined from mass spectrometry measurements at 2 mm from the plasma torch in central position (Ar 30 L/min, $O_2$ 0 mL/min, 80 W)*

Mass spectrometry measurements (cf. figure 6) reveal that, in a central position under the plasma torch, the plasma afterglow creates a protective atmosphere for distances smaller than 6 mm. For greater distances, $N_2$ and $O_2$ molecular concentrations rapidly increase. Accordingly, emission from the 2$^{nd}$ positive system of $N_2$ is detected in the same range of distances (cf. figure 7). The decrease in emission intensity from 9 mm onwards must be related to the decaying Ar excitation temperature.

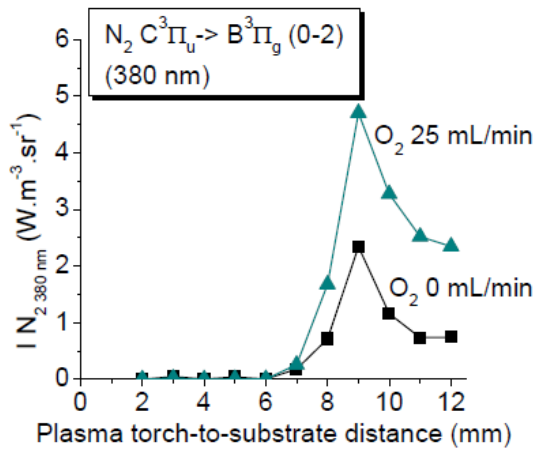

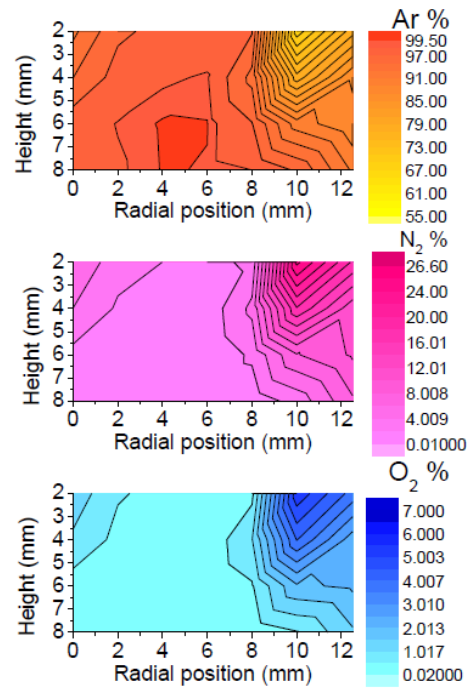

*Figure 7. $N_2$ $C^3\Pi_u \rightarrow B^3\Pi_g$ (0-2) band emission intensity versus plasma torch-to-substrate distance (Ar 30 L/min, O2 0 mL/min 80 W)*

*Figure 8 Ar, $N_2$ and $O_2$ relative concentrations determined from mass spectrometry measurements versus height and axial position in the plasma afterglow (Ar 30 L/min, $O_2$ 0 mL/min 80 W) for a plasma torch-to-substrate distance of 9 mm*

Figure 8 presents Ar, $O_2$ and $N_2$ relative concentrations obtained from mass spectrometry measurements ($H_2O$ and $CO_2$ molecules were included in the total composition) versus height and axial position from the center of the plasma torch for a fixed distance of 9 mm. Gradients in species concentrations are observed, the N2 concentration reaching 26 % close to the edge of the plasma torch at a height of 2 mm.

## 4. Conclusions

The influence of the $O_2$ gas flow rate and the source power on the properties (gas temperature, Ar excitation temperature, relative concentrations of O atoms and OH radicals) of an Ar/$O_2$ plasma afterglow have been evaluated. The OH concentration is seen to decrease with an increase in plasma source power and is higher when $O_2$ is injected. The Ar excitation temperature is seen to increase with $O_2$ gas flow when greater than 5 mL/min, and increase with source power. Optical emission spectroscopy and mass spectrometry measurements show that for plasma torch-to-substrate distances lower than 6 mm, the afterglow creates a protective atmosphere. When the distance increases, the Ar excitation temperature decreases, and $N_2$ and $O_2$ molecular concentrations rapidly increase. Gradients in Ar, $O_2$ and $N_2$ concentrations are then observed.

## 5. Acknowledgments

This work was carried out in the framework of the Interuniversitary Attraction Pole program "Plasma Surface Interactions" financially supported by the Belgian Federal Office for Science Policy (BELSPO).